\documentclass[reprint,amsmath,amssymb,english,aps,pra]{revtex4-1}
\usepackage[T1]{fontenc}
\usepackage{amsmath}
\usepackage{amssymb}
\usepackage{graphicx}
\usepackage{babel}
\usepackage{color}
\usepackage{braket}
\usepackage[normalem]{ulem}
\usepackage[colorlinks,bookmarks=false,citecolor=magenta,linkcolor=blue,urlcolor=black]{hyperref}

\usepackage{xcolor}

\newcommand{\be}{\begin{equation}}
\newcommand{\ee}{\end{equation}}
\newcommand{\bi}{\begin{itemize}}
\newcommand{\ei}{\end{itemize}}

\begin{document}

\title{Time dynamics with matrix product states: \\
Many-body localization transition {of} large  systems revisited}
\author{Titas Chanda$^1$,
  Piotr Sierant$^1$,} 
 \email{piotr.sierant@uj.edu.pl}
   \author{Jakub Zakrzewski$^{1,2}$}

\affiliation{\mbox{$^1$Instytut Fizyki imienia Mariana Smoluchowskiego, Uniwersytet Jagiello\'nski, \L{}ojasiewicza 11, 30-348 Krak\'ow, Poland}\hfill
\mbox{$^2$Mark Kac Complex Systems Research Center, Jagiellonian University, \L{}ojasiewicza 11, 30-348 Krak\'ow, Poland}
}
\date{\today}% It is always \today, today,
             %  but any date may be explicitly specified
\begin{abstract}
We compare accuracy of two prime {time evolution algorithms} involving Matrix Product States - tDMRG (time-dependent density matrix renormalization group) and TDVP (time-dependent variational principle). The latter is supposed to be superior within a limited and fixed auxiliary space dimension. Surprisingly, we find that the performance of algorithms depends on the model considered. In particular, many-body localized systems as well as the crossover regions between localized and delocalized phases are better described by tDMRG, contrary to the delocalized regime where  TDVP indeed outperforms tDMRG in terms of accuracy and reliability. As an example, we study many-body localization transition in a large size
Heisenberg chain. We discuss drawbacks of previous estimates [Phys. Rev. B{\bf 98}, 174202 (2018)] of the critical disorder strength for large systems. 
\end{abstract}
\maketitle

\section{Introduction}
In recent years, matrix product states (MPS) \cite{Schollwoeck11, Orus14}  {have} become  major tools in representing quantum states and their dynamics in many-body one-dimensional systems.
For static calculations, the MPS-based density matrix renormalization group (DMRG) \cite{White92, White92, White05, Hubig15} has become the \emph{de facto} standard for obtaining the ground and a few low-lying excited states of many-body Hamiltonians because of the accuracy and reliability of the method. 
In case of  time-evolution, the breakthrough came with the development of time-evolving-block-decimation (TEBD) algorithm \cite{Vidal03,Vidal04} and its variant, time-dependent density matrix renormalization group (tDMRG) \cite{White92,White93,White04}, with similarities and differences 
between the two approaches being discussed soon \cite{Daley04}. The simple and transparent algebraic properties of MPS used in the algorithms shortly led to different developments, such as treatment of translationally invariant infinite systems \cite{Vidal07,Orus08}, matrix product operators (MPO) technique \cite{Zaletel15}, time-dependent variational principle (TDVP) scheme with its one-site and two-site versions \cite{Haegeman11, Koffel12, Haegeman16} to name a few. Rapid development in the field has been summarized in seminal reviews \cite{Schollwoeck11, Paeckel19} with leading methods presented and compared in accuracy and speed on representative examples (see also \cite{Leviatan17, Kloss18, Goto19, Hemery19}). 
The cases studied correspond to very demanding examples with a rapid
entanglement growth (for initial weakly entangled states), 
which show that the TDVP (with two-site version in the initial phases and one-site version afterwards) is less error-prone and more accurate than other available methods, and should be selected as a method of choice. This seems understandable as the variational approach should, in most cases, lead to optimal solutions. 

On the other hand, time-dynamics of large systems have been often addressed with  TEBD-based approaches \cite{Clark04,Kollath06,Kollath07,Znidaric08,Wall09,Zakrzewski09,Lacki12,Lacki13,Delande13,Karrasch14, Huang17}, and the
MPS-based techniques seem also a natural tool to consider time-dynamics in many-body localization (MBL) problems, especially as experiments \cite{Schreiber15} deal with systems not amenable to exact diagonalization techniques \cite{Znidaric16,Prelovsek16,van_Nieuwenburg17,Sierant17,Sierant17b,Sierant18,Zakrzewski18}. Recent study using TDVP in XXZ spin chain addresses the MBL transition in detail \cite{Doggen18} claiming that the critical disorder corresponding to MBL transition strongly depends on the system size. We also analyse this issue in this work, but first we compare the performance of tDMRG and TDVP in systems close to localized regime. Surprisingly, we have found indications that when non-ergodic features in the dynamics become pronounced (as at the onset of MBL regime) the standard, old-fashioned tDMRG may perform very well. Moreover, results obtained with too small auxiliary space lead to spurious effects for TDVP that may affect the conclusions concerning intermediate and long-time dynamics.
We believe this interesting observation calls for a detailed studies presented below.  Having an excellent overview of different methods at hand \cite{Paeckel19}, we only briefly mention our implementations of algorithms in Section~\ref{sec:tools}. The core of our results is presented in Section~\ref{sec:res} where we compare the performance of TDVP based and TEBD-based schemes with a numerically exact
propagation obtained via Chebyshev expansion of the evolution operator \cite{Weisse06}. Larger system sizes, where no exact results are available, are discussed in Section~\ref{sec:large}, while in Section~\ref{sec:crit} we discuss the difficulties with an estimation of the critical disorder value for disordered Heisenberg spin chain from such time-dynamics. We provide a state-of-art estimate for the critical disorder value for large system sizes. To some extent, they confirm the existence of MBL transition for large systems, contrary to recent analysis \cite{Suntais19} that questions the MBL existence in the thermodynamic limit.
We conclude  in Section~\ref{sec:outlooks}. 

\section{Numerical tools}
\label{sec:tools}
In this section, we briefly discuss our implementations of two MPS-based time-evolution strategies used in this work -- A:  time-dependent density matrix renormalization group (tDMRG) \cite{White92,White93,White04, Daley04}, as a variant of time evolving block decimation (TEBD) technique  \cite{Vidal03,Vidal04, Paeckel19, Schollwoeck11} and B: recently developed time-dependent variational principle (TDVP) \cite{Haegeman11, Koffel12, Haegeman16}. 

\subsection{tDMRG}
Instead of using more commonly used Suzuki-Trotter decomposition for implementing tDMRG/TEBD methods,
we use the second-order Sornborger-Stewart \cite{Sornborger99} decomposition of the time-evolution operator as first described in \cite{White04}.  For a nearest-neighbor many-body Hamiltonian 
$H = \sum_{i=1}^{L-1} H_i$ consisting of $L$ sites,
where $H_i$ denotes the term on $i^{th}$ bond, the decomposition of the small-time-evolution operator is given by
\begin{equation}
e^{-i H \delta t} \approx e^{-i H_1 \frac{\delta t}{2}}  \ldots e^{-i H_{L-1}\frac{\delta t}{2}} e^{-i H_{L-1} \frac{\delta t}{2}} \ldots e^{-i H_{1} \frac{\delta t}{2}},
\label{eq:ssdecom}
\end{equation}
which incurs error of the order $O(\delta t^3)$  as similar to the second-order Suzuki-Trotter one. The term in Eq.~\eqref{eq:ssdecom} is then applied to a physical state,  represented by MPS ansatz, in left-to-right and right-to-left sweeps exactly as in a two-site DMRG algorithm \cite{White04, Daley04}.  After each such a sweep, the dimension of the auxiliary space (MPS bond dimension) grows with time, and if the bond dimension becomes larger than a desired value, say $\chi$, it is truncated by keeping only $\chi$ largest singular values after a renormalization. In our simulations, we set the time-step for tDMRG as $\delta t =0.02$ in units of the Hamiltonian. 

\subsection{TDVP}

In TDVP, the time-evolution of a MPS is attained by computing the action of the Hamiltonian only along the tangent direction to  the present variational MPS manifold, describe{d} by the MPS bond dimension $\chi$. Mathematically, instead of solving the time-dependent Schr{\"o}dinger equation, we solve
\begin{equation}
\frac{d \ket{\psi(t)}}{d t} = - i \ \mathcal{P}_{MPS_{\chi}}  \ H \ket{\psi(t)},
\end{equation}
where the projector $\mathcal{P}_{MPS_{\chi}}$ projects the action of the Hamiltonian $H$ to the tangent plane of the variational MPS manifold of bond dimension $\chi$.
In our work, we follow the prescription described in Refs. \cite{Haegeman16, Paeckel19} (see Ref. \cite{Koffel12} for an alternative one) to implement both two-site and one-site versions of TDVP. Since the two-site version allows to grow the MPS bond dimension dynamically during the TDVP sweeps, we employ this version in the initial  stage of the dynamics. As soon as the bond dimensions in the bulk of the MPS is saturated to a desired value, say $\chi$, we switch to the one-site version, where dimensions of auxiliary spaces do not change. Such a hybrid method of time-evolution using TDVP has been argued to incur lesser errors \cite{Paeckel19, Goto19}. We use the step-size $\delta t = 0.1$ (in units of the Hamiltonian) for TDVP calculations, however, since we use properly converged Lanczos exponentiation \cite{Hochbruck97} in TDVP simulations, smaller step-sizes do not affect the results indicating the convergence with respect to the time step (see Fig.~\ref{timeaccuracy} below). 

\section{MPS time evolution versus ``exact'' results}
\label{sec:res}

As a model to study, we take the Heisenberg spin chain  with random field, often studied in the context of MBL \cite{Mondaini15,Luitz15}. Explicitly we consider
the Hamiltonian
\begin{equation}
	H = J\sum_{i=1}^{L-1}\vec S_{i} \cdot \vec S_{i+1}+
\sum_{i=1}^{L}h_{i}S_{i}^{z}, 
	\label{hamXXZ}
\end{equation}
where  $\vec{S_i}$ are spin-$1/2$ degrees of freedom at site $i$,  $J$ is the spin coupling
 strength, from now on we set $J\equiv 1$ fixing the energy scale and
$h_{i}$  is the  random magnetic field drawn in our examples form
 a random uniform distribution in $\left[-W; W \right]$ 
with $W$ denoting the disorder strength. The Hamiltonian (\ref{hamXXZ}) maps 
directly {to an} interacting spinless fermions model making it, in principle, accessible in cold atoms experiment.
Level statistics analysis of small systems ($L\leq 22$) indicates a transition between extended  and  MBL regimes at {$W_c = 3.72(6)$} 
(the value extrapolated {to} the thermodynamic limit  via a finite size analysis of level spacing ratios) \cite{Luitz15}.  Similar value is obtained from
the entanglement entropy scaling \cite{Luitz15}. On the other hand, the recent study \cite{Doggen18} of larger systems (size $L=100$)
gives the estimate of the transition at a much larger value $W_c\approx 5.5$ on the basis of time dynamics obtained using TDVP.
This obviously contradicts the {above} mentioned result based on finite size scaling of data for systems with $L\leq 22$ and suggests that either the time-dynamics 
of observables gives different answer than {analysis of level statistics
and properties of eigenvectors or results
obtained by TDVP \cite{Doggen18} have to be reconsidered.

We analyse a time-evolution obtained for the Hamiltonian \eqref{hamXXZ} starting from the  N\'eel state with every second spin pointing up and every second spin down
{$|\psi\rangle = | \uparrow \downarrow \ldots \uparrow \downarrow\rangle$}. In the fermionic language such a state corresponds to a perfect density wave with every second site occupied. It is known that the properties of the system \eqref{hamXXZ} may depend on energy, leading to the so called \emph{mobility edge} for the precise location of the MBL transition \cite{Luitz15}, 
such that 
 to probe the system properties initial states should be adjusted to a particular realization of disorder to assure probing of the same energy region. We nevertheless choose a single
state mentioned above to facilitate comparison with \cite{Doggen18}. It is worth remembering that a similar state was also prepared in the 
experiments \cite{Schreiber15} to study MBL dynamics.  
In all numerical examples shown, we disregard two sites on each edge of the chain to minimize
the effect of open boundary conditions. This allows us for a better comparison of systems of different sizes as discussed in Section~\ref{sec:crit}.

Let us commence our numerical studies with moderate size system $L=26$ with open boundary conditions within total $S_z=0$ sector (corresponding to fermionic half filling). Its exact evolution using standard Hamiltonian diagonalization approach would be a formidable task. {On the contrary,} Chebyshev polynomial expansion of the time-evolution operator \cite{Weisse06} {used already a few times in the context of MBL
\cite{Bera17,Sierant19a,Weiner19}} allows us to obtain
numerically exact time-evolution up to quite long times. Those results serve as  
benchmarks against which we compare the performance of tDMRG and TDVP.  
For these studies, we consider the same 200 disorder realizations for both methods and
monitor the time dependence of the imbalance, i.e., the normalized difference between 
magnetizations of even and odd lattice sites
\begin{equation}
 I(t) = C  \sum_{i=1}^L (-1)^i  \langle \psi(t) |S^z_i | \psi(t)\rangle, 
\end{equation}
where $| \psi(t)\rangle = e^{-i H t} |\psi\rangle$ and the constant $C$ assures that $I(0) = 1$
for the initial N\'eel state. As in experiment \cite{Schreiber15}, a decay of the imbalance with time indicates a delocalized regime  
while its saturation at some finite value points towards MBL. 

\begin{figure}
  \includegraphics[width=.9\columnwidth]{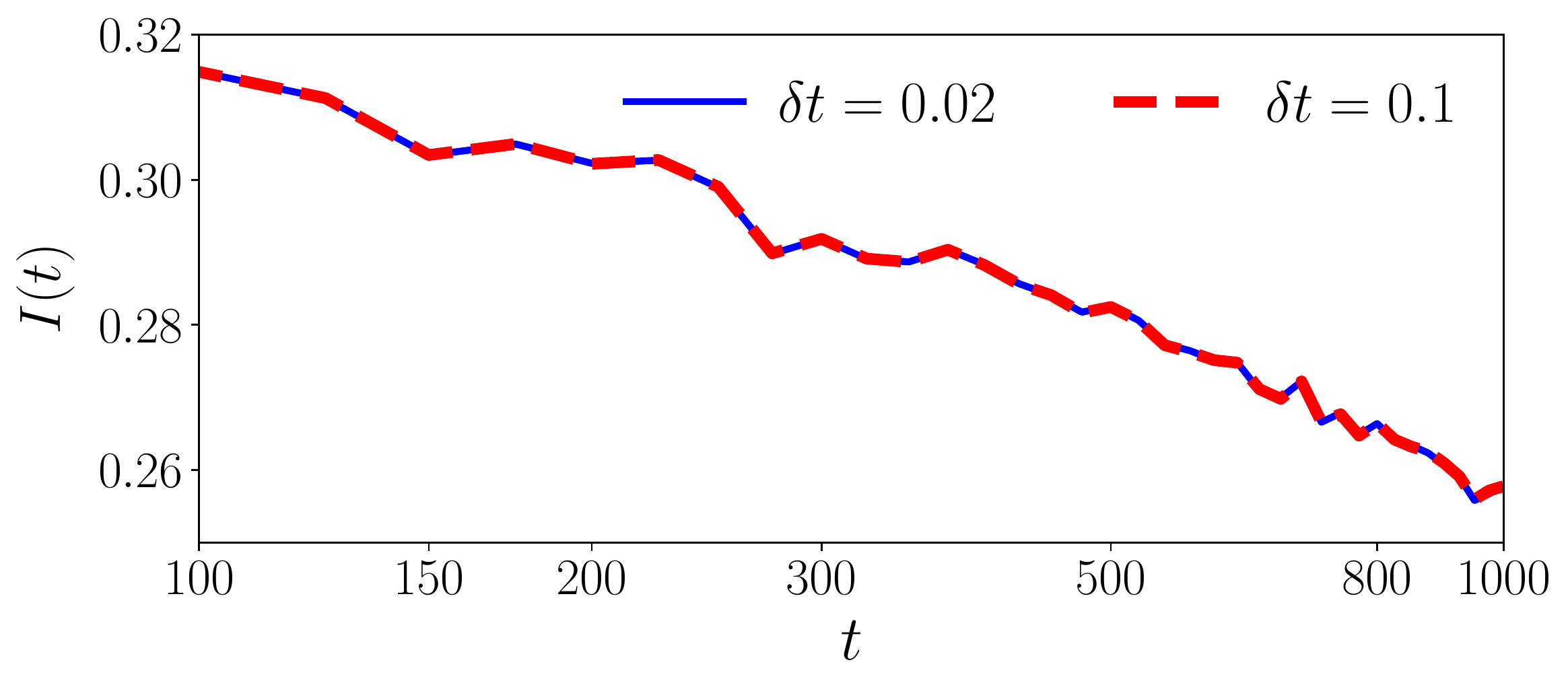}
\caption{Imbalance $I(t)$ as a function of time in the delocalized regime $W=3$ for the system-size $L=26$ obtained with TDVP for $\chi=256$ and 200 disorder realizations for two different time steps as indicated in the figure. Only $t = 25n$, $n$ being a positive integer, points are plotted for clarity. The differences in the disordered averaged imbalance are of the order of $10^{-4}$ indicating a full convergence with the time step taken.}
 \label{timeaccuracy}
\end{figure}

\begin{figure}
  \includegraphics[width=\columnwidth]{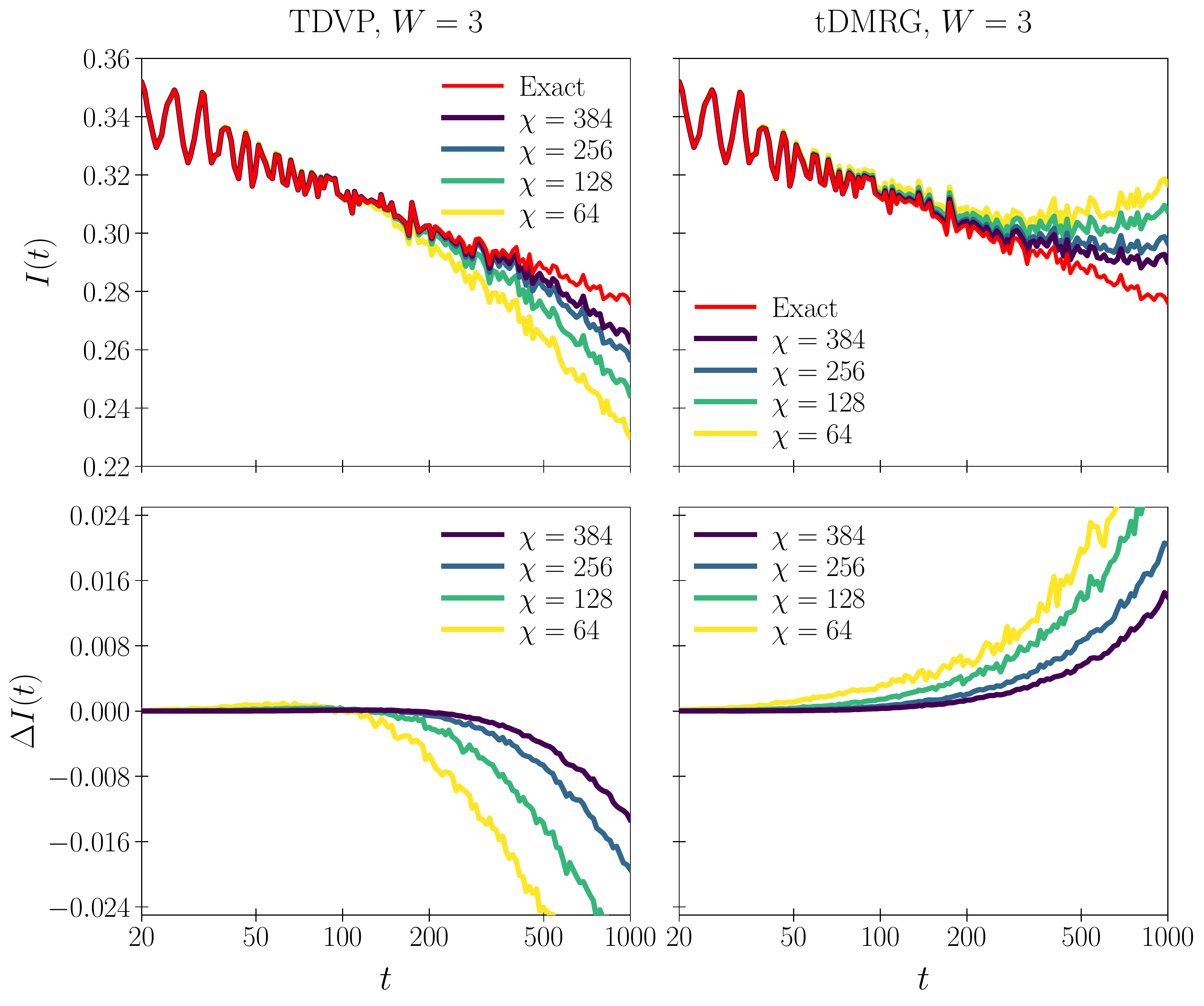}
  \caption{Imbalance $I(t)$ as a function of time in the delocalized regime, $W=3$, for $L=26$ spin chain for different dimensions ($\chi$) of the auxiliary space as indicated in the legend. Here we take same 200 disorder realizations for all the techniques.
Note that  TDVP consistently leads to faster than exact decay (left), while tDMRG (right) reveals a false saturation for too small $\chi$. The magnitude of the error, $\Delta I(t)$, for the same $\chi$ is typically smaller for TDVP as shown in the bottom row where deviations from the exact results are plotted.}
 \label{fig2:ext}
\end{figure}

Let us now consider how the exact time-dynamics of imbalance is approximated with TDVP and tDMRG algorithms. Fig.~\ref{fig2:ext} shows the exemplary time evolution of imbalance $I(t)$ for $W=3$ and $L=26$ for different dimensions $\chi$ of the auxiliary space as compared with the ``exact'' result obtained by the Chebyshev approach. Interestingly, TDVP leads to a much faster decay of the imbalance than the exact result, approaching
the ``exact'' data from below {as the bond dimension $\chi$ is increased}. tDMRG does just the opposite - too small $\chi$ values lead to a false saturation of the imbalance. 
Errors {$\Delta I(t)$} {with respect} to the exact result are comparable being a bit smaller for TDVP. 

\begin{figure}
  \includegraphics[width=\columnwidth]{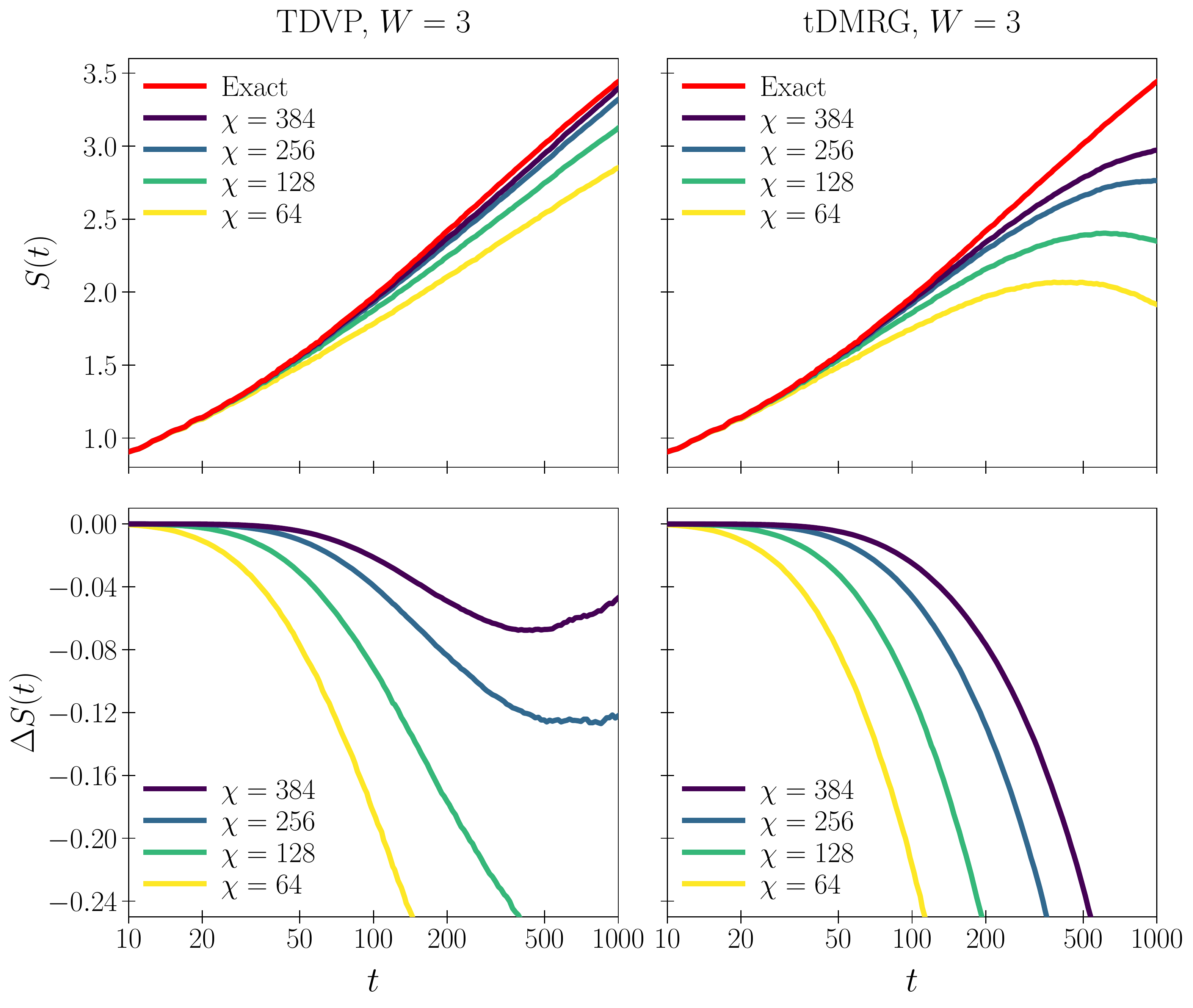}
  \caption{Entanglement entropy growth in time in the middle of the $L=26$ chain for $W=3$.  Note that consistently TDVP (top left) follows the exact growth for slightly longer period while tDMRG (top right) deviates faster showing a stronger false saturation. The errors in both cases are shown in the bottom row.}
 \label{fig3:entro}
 \end{figure}
The superiority of TDVP approach is{, however, }
 clearly visible when the entanglement entropy $S(t)$ in the 
middle of the chain is considered. We trace out half of the chain (13 sites in this case) and plot the entanglement entropy growth as a function of time in Fig.~\ref{fig3:entro}.  While both TDVP and tDMRG underestimate the entanglement entropy growth, TDVP follows the exact
growth for a slightly longer time, while tDMRG yields a spurious decrease of the entanglement entropy at large times.

\begin{figure}
  \includegraphics[width=\columnwidth]{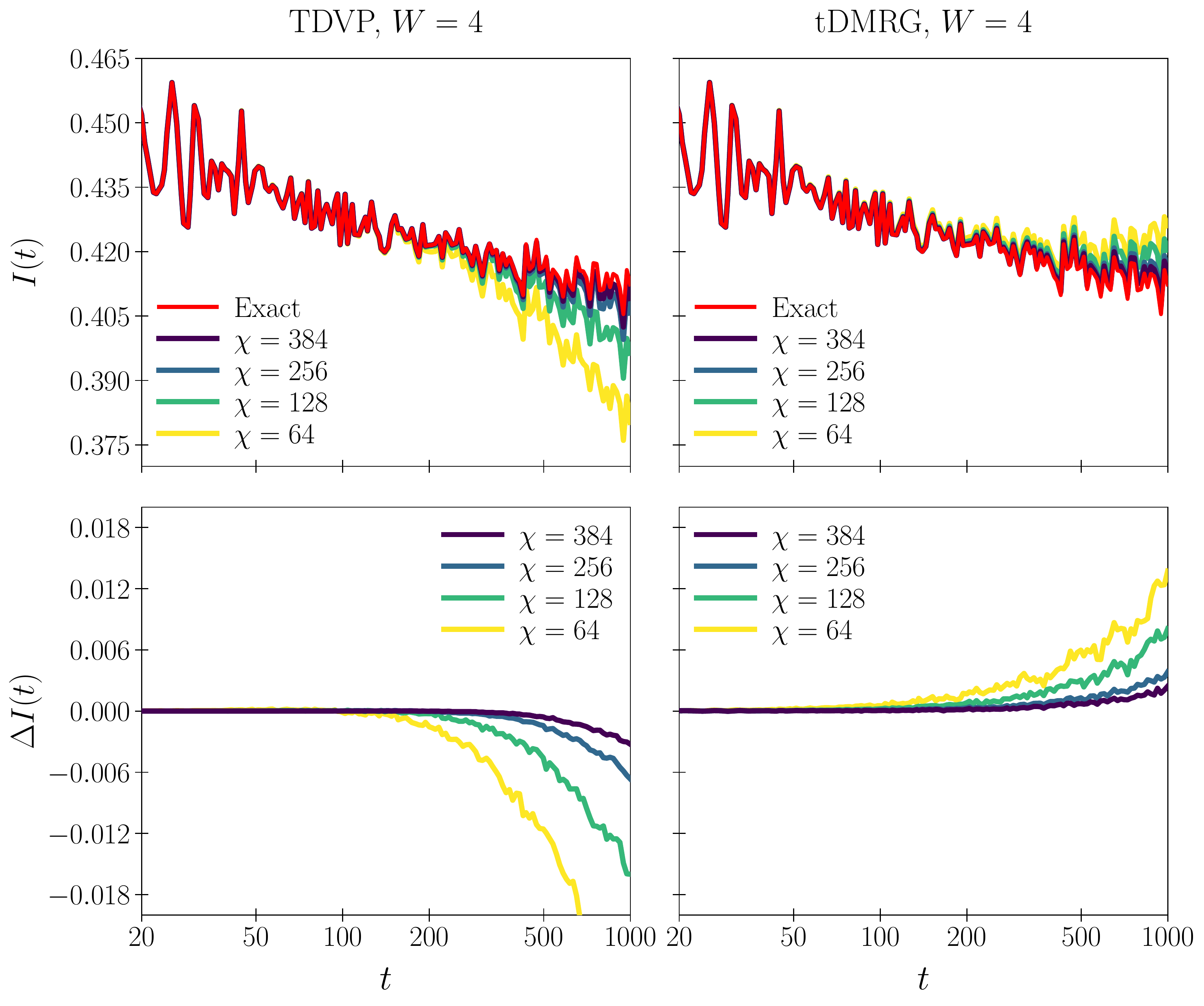}
  \caption{Same as in Fig.~\ref{fig2:ext} but for $W=4$, where top panel shows the dynamics of imbalance for different values of bond dimension for both TDVP (top left) and tDMRG (top right),
 and the bottom panel shows the corresponding errors with respect to the exact Chebyshev result (TDVP on bottom left, tDMRG on bottom right). Observe that tDMRG becomes, surprisingly, more accurate with increasing disorder.}
 \label{fig4:imba}
 \end{figure}
Let us now come closer to MBL crossover region by increasing the disorder amplitude to $W=4$. This disorder value is higher than $W_c=3.7$, the critical disorder value for the MBL transition extracted from the finite size analysis  in \cite{Luitz15}. 
The direct inspection of the numerical data as well as  of the errors with respect to exact results indicates that, surprisingly,  tDMRG becomes superior to TDVP showing consistently smaller errors with respect to Chebyshev propagation exact results (see Fig. \ref{fig4:imba}).

\begin{figure}
  \includegraphics[width=\columnwidth]{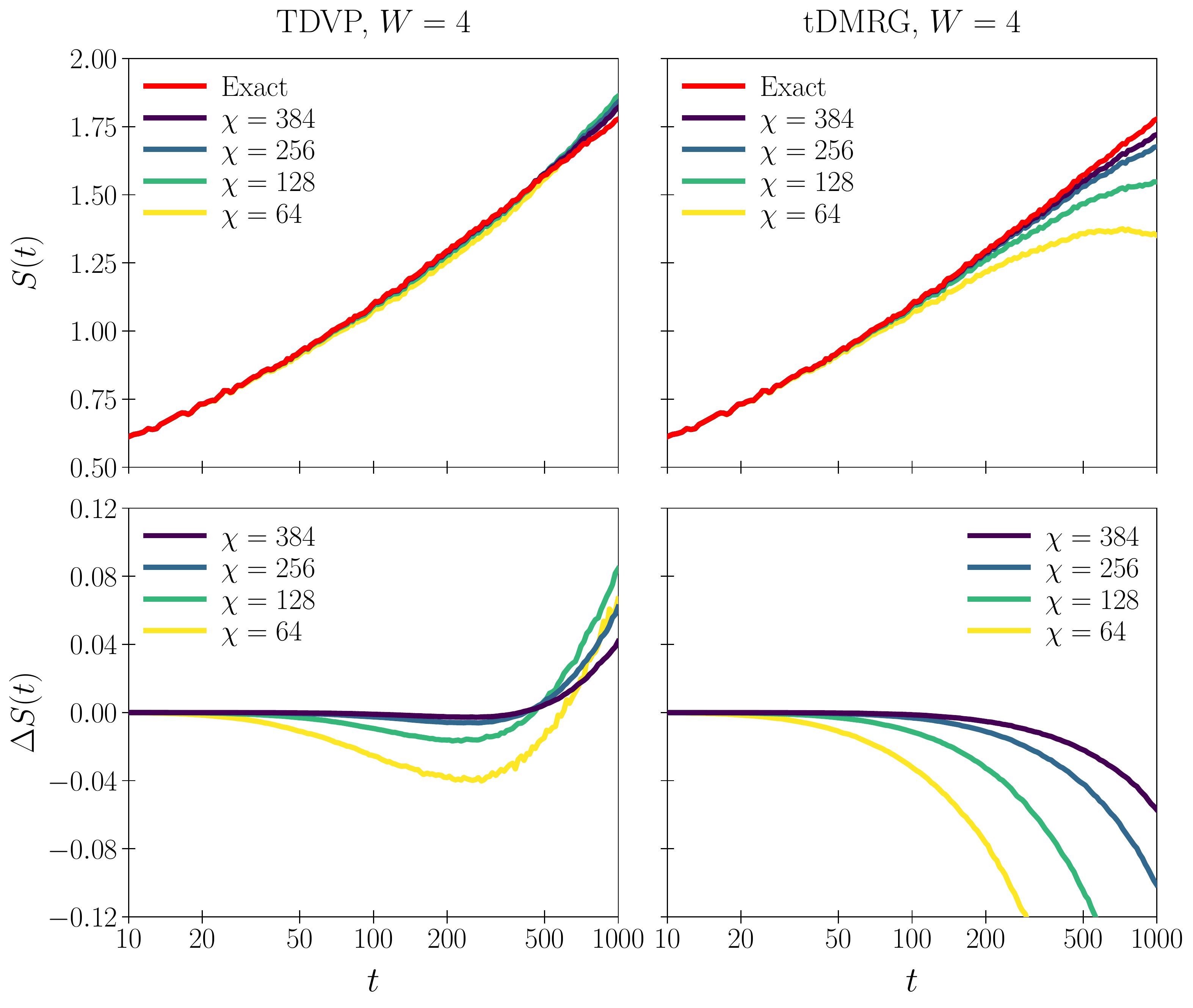}
  \caption{The entanglement entropy growth for $L=26$ and $W=4$ as compared to exact results due to Chebyshev propagation (top) and the residuals (bottom). Observe that TDVP overshots the growth for large times when $\chi$ is not sufficient for convergence. The behavior of tDMRG is more predictible converging smoothly with increasing $\chi$.}
 \label{fig4:ent}
 \end{figure}
The similar conclusion can be reached by analyzing the entanglement entropy (Fig. \ref{fig4:ent}), where tDMRG shows a consistent behavior, namely, as long as $\chi$ is sufficient to represent the dynamics, it gives a correct prediction for the entanglement entropy, and for longer times it always
underestimates the exact value. The behavior of TDVP calculation at long times, on the contrary, is markedly different. For longer times, when the bond dimension $\chi$  is insufficient to represent the dynamics, TDVP overshoots the entropy growth. In effect, at a given value of $\chi$ the entropy coming from TDVP may overestimate the exact value, a comparison of values obtained with TDVP for two different $\chi$ values might be of little help. For instance, increasing the bond dimension from $\chi=64$ to $\chi=128$ one could conclude that entanglement entropy growth is quicker than logarithmic at times $t > 300$. Only once the 
bond dimension is sufficiently large, $S(t)$ obtained from TDVP converges monotonically to the exact result.

By comparison, the entropy {$S(t)$} in tDMRG underestimates the exact result
in a predictable, monotonic way. Therefore, our results indicate that the
value of $S(t)$ obtained from tDMRG can be regarded a lower bound for the
entanglement entropy for any time.

We have presented the data for $L=26$, the data for $L=20$ show a similar behavior with tDMRG becoming a method of choice for even lower values of the disorder (as the ``critical'' region occupies larger interval of disorder amplitudes for a smaller system).

Some understanding of the different convergence of both methods may be obtained considering different ways in which necessary approximations are made. In tDMRG, when the wavevector spreads over the allowed Hilbert space in auxiliary dimension, we move to a higher MPS manifold (bigger auxiliary space)
in a given step. Then we come back to the assumed size (as determined by $\chi$) by truncation and renormalization of relevant singular values. In this process contributions leading to the higher auxiliary spaces are removed, and ``less entangled'' and more localized components of the wavepacket are relatively better reproduced, as the relevant singular values are much more skewed towards higher values. In the same spirit, it was shown in a Bose-Hubbard case study \cite{Lacki12} that a long time (beyond any reasonable time scale for a full convergence) evolution in tDMRG allows one to extract localized excited eigenstates. On the other hand, in one-site TDVP (which is used when the maximum allowed size of the auxiliary space is achieved), we move within the same MPS manifold by only considering the action of the Hamiltonian projected into the tangent space of the MPS manifold, and the source of error comes from such a projection.
In a typical situation, for clean or delocalized systems, this strategy also leads to eventual underestimation of entanglement entropy. However,  the outcome is much different for strongly disordered systems, as the unconverged data look spuriously ``more delocalized'', and surprisingly,
the entanglement entropy is overestimated at long times if $\chi$ is not sufficient to describe the dynamics. This overestimation is counter-intuitive as smaller values of $\chi$ should produce lower entanglement entropy. Such observations can motivate further studies regarding time-evolution algorithms using tensor-networks in localized systems.

\section{tDMRG versus TDVP for large systems}
\label{sec:large}

As a standard ``large'' system we consider the same model (as given in Eq.~\eqref{hamXXZ}) with $L=50$ sites with open boundary conditions. This is a typical system size met in cold atoms experiments \cite{Schreiber15}. The previous TDVP based analysis \cite{Doggen18} has concentrated on $L=100$ showing that $L=100$ and $L=50$ do not differ substantially.  Still $L=50$ is computationally less expensive and, on the other hand, it seems sufficiently large to ensure that boundary effects are small. We consider 200 realizations of disorder  in our analysis.

\begin{figure}
  \includegraphics[width=0.9\columnwidth]{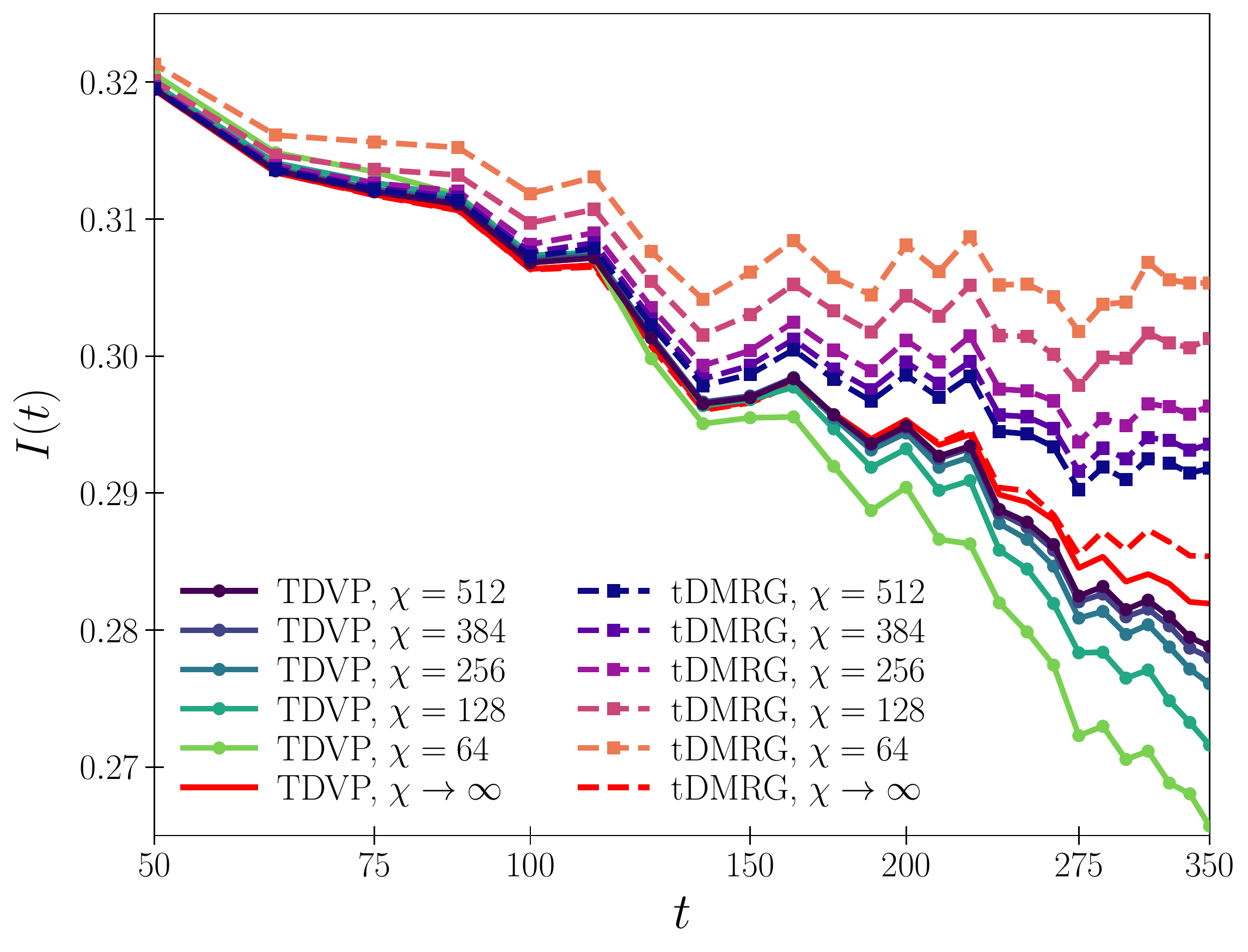}
  \caption{Time evolution of the imbalance for disorder strength $W=3$ (on the delocalized side) for different dimensions of the auxiliary Hilbert space $\chi$ as indicated in the figure. For this disorder value even $\chi=512$ is not sufficient for the convergence up to times $t=350$.  Top five curves correspond to tDMRG showing an apparent saturation of the imbalance, diminishing with increasing $\chi$.  Five bottom curves correspond to TDVP with an opposite behavior, the bigger $\chi$ the slower the decay. Red lines without symbols correspond to extrapolations to $\chi\rightarrow\infty$ limit -- see text. The same 200 disorder realizations are used for each curve.}
 \label{fig50imb}
 \end{figure}

Let us consider again the $W=3$ case first. For this disorder value and due to the large size, we
expect the system to be on the delocalized side of the transition with TDVP working significantly better than tDMRG. The results of the simulations for the dynamics of imbalance $I(t)$ using both TDVP and tDMRG are shown in Fig.~\ref{fig50imb} again for the antiferromagnetic N{\'e}el initial state. Firstly, we observe a similar behavior as for smaller system sizes. Too small $\chi$ leads to a false saturation of the imbalance for tDMRG, with increasing $\chi$
the imbalance decays for longer times. TDVP shows an opposite effect, too fast decay for small $\chi$. The ``exact'' result is expected 
somewhere between the tDMRG and TDVP curves. Observe that TDVP seems to be almost converged when comparing $\chi=384$ and $\chi=512$ curves while the corresponding tDMRG curves show a bigger discrepancy. Both methods should converge to the same result, but looking at the curves it seems apparent that a much larger bond dimension than used in the simulations is needed to sufficiently explore the Hilbert space and get a truly converged result even for TDVP. 

\begin{figure}
  \includegraphics[width=0.9\columnwidth]{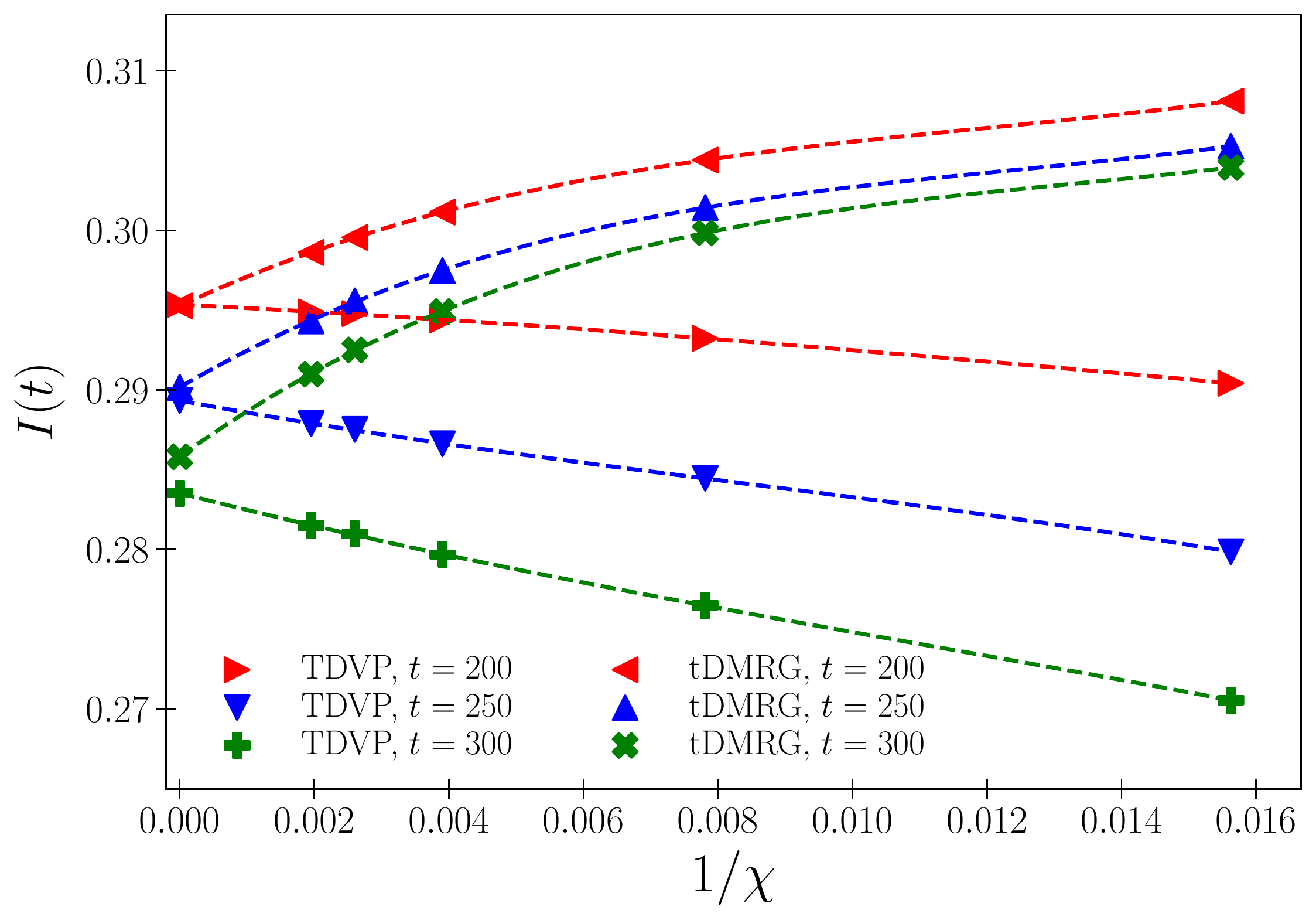}
  \caption{Extrapolation of the imbalance for $W=3$ to $\chi\rightarrow\infty$ by fitting 
  a third-order polynomial of $1/\chi$ to the data obtained for selected times and same values of $\chi$ as presented in Fig.~\ref{fig50imb}. Fitting errors are less than $10^{-10}$.} 
 \label{fig50imb_fit3}
 \end{figure}
This aim may be reached by making an analysis of result obtained for different $\chi$
in an attempt to extrapolate the results in the limit $\chi\rightarrow \infty$. Inspired by standard finite-size extrapolation, we present the data for chosen instances of time as a function of $1/\chi$ (see Fig.~\ref{fig50imb_fit3}). Fitting a simple third-order polynomial leads to a surprisingly nice agreement between the two approaches. For longer times the agreement deteriorates as results for tDMRG and TDVP differ too much for simple extrapolation procedure to hold. The results for such extrapolations are shown in Fig.~\ref{fig50imb} as red lines without symbols.
 
\begin{figure}
  \includegraphics[width=\columnwidth]{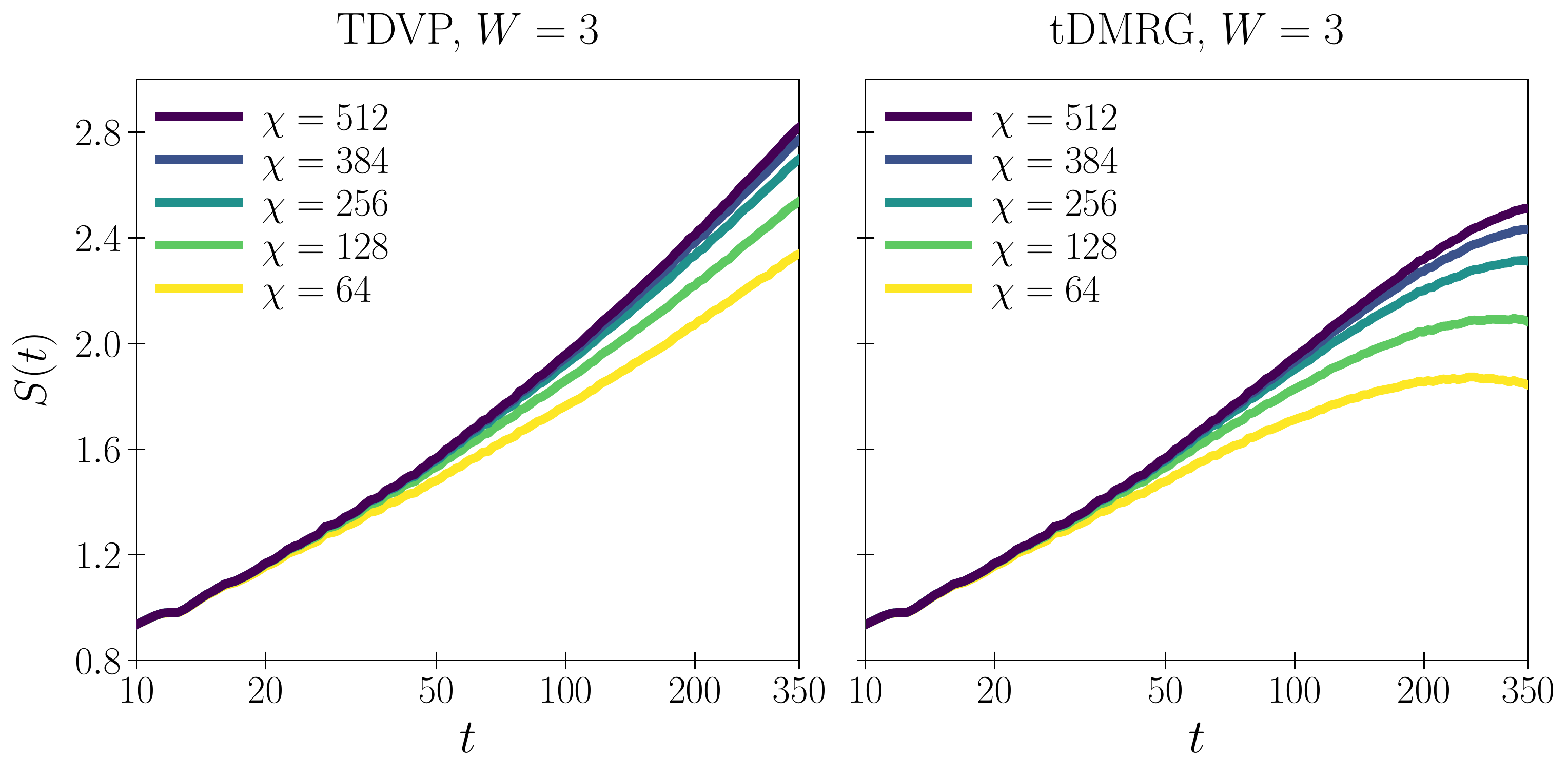}
  \caption{The entanglement entropy growth in the middle of the chain for $L=50$ and $W=3$  corresponding to the imbalance depicted in Fig.~\ref{fig50imb} for TDVP (top panel) and tDMRG(bottom panel). For entanglement entropy,  increase in $\chi$ lead to higher values, and
  the differences between both numerical approaches are significant with TDVP seemingly being better converged. }
 \label{fig50ent}
 \end{figure}

The corresponding growth of entanglement  entropy $S(t)$ in the middle of the chain (computed after tracing out half of the chain)  in time is depicted in Fig.~\ref{fig50ent}. Observe a clear 
saturation of entropy for tDMRG when the chosen $\chi$ values are insufficient to represent the dynamics, and  the almost converged entropy growth for TDVP,
where $\chi=512$ data is reasonably fitted with a power law growth $S(t) \propto t^{0.29}$  in accordance with the expectation that the entanglement entropy growth is faster than logarithmic 
on the delocalized side of the crossover. 

\begin{figure}
  \includegraphics[width=0.9\columnwidth]{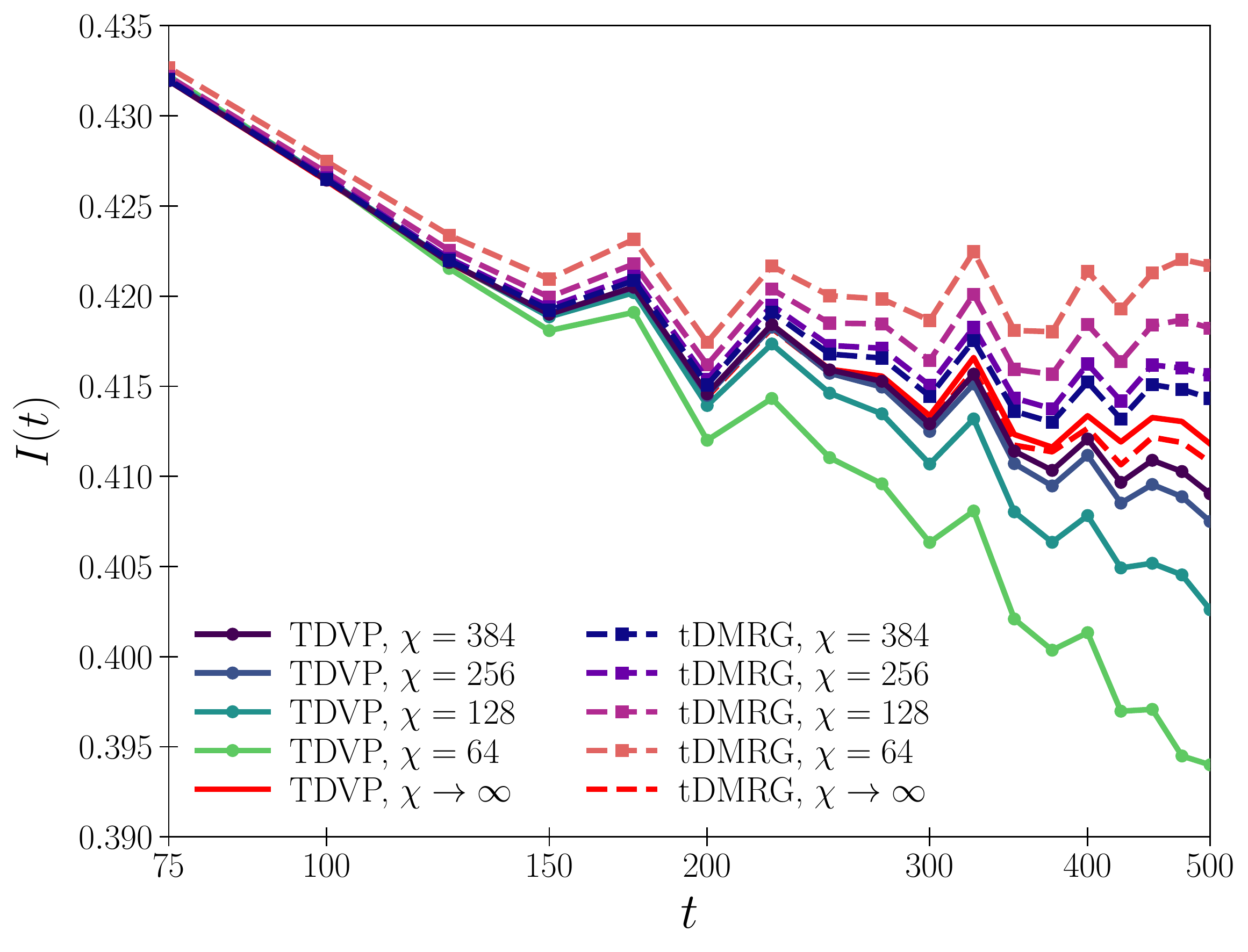}
  \caption{Time dependence of the imbalance for $W=4$ and different values of $\chi$. Top four curves correspond to tDMRG calculation, while bottom curves for TDVP as indicated in the figure. As in Fig.~\ref{fig50imb} curves without symbols represent the result of $\chi\rightarrow\infty$ extrapolation.}
 \label{fig50_4imb}
 \end{figure}

 Let us consider slightly bigger disorder $W=4$. The corresponding evolution of the imbalance is shown in Fig.~\ref{fig50_4imb} for different $\chi$ values both for TDVP and tDMRG using  200 realizations of disorder. While differences between various $\chi$ values indicate smaller deviations in imbalance for tDMRG, 
 TDVP {with $\chi=384$} matches almost exactly TDVP result for $\chi=512$ up to $t=350$.
 
 \begin{figure}
  \includegraphics[width=0.9\columnwidth]{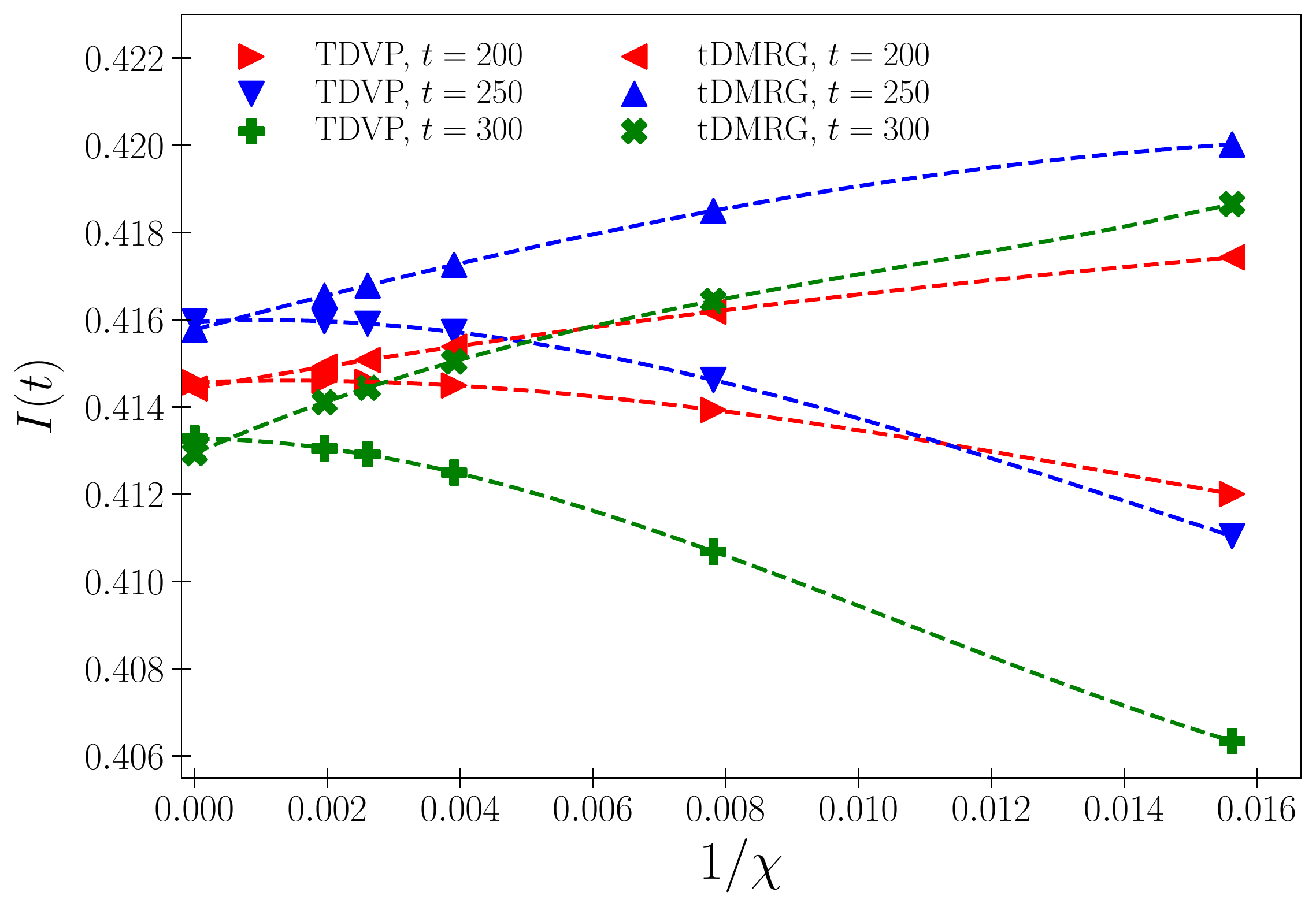}
  \caption{
Extrapolation of the imbalance for $W=4$ to $\chi\rightarrow\infty$ by fitting 
  a third-order polynomial of $1/\chi$ to the data obtained for selected times and different values of $\chi \in [64, 128, 256, 384, 512]$. Fitting errors are less than $10^{-10}$.}
 \label{fig50imb_fit4}
 \end{figure}

As in the case of $W=3$, we may attempt extrapolation to larger $\chi$ values -- see  Fig.~\ref{fig50imb_fit4} for examples of the procedure while the result of the extrapolation of time dependence are presented in Fig.~\ref{fig50_4imb}. Again the agreement is quite nice. Observe a non-monotonous decrease of estimated $I(t)$ with time, which reflects time fluctuations of imbalance
 -- compare Fig.~\ref{fig50_4imb} -- related to a small number of disorder realizations. 
 The corresponding entropy growth, this time relative to $\chi=512$, data is represented in Fig.~\ref{fig50_4ent}. Although the entanglement entropy error is much smaller for TDVP showing an apparent better convergence,  the curves are not monotonically decreasing with time which strongly suggests that the entropy obtained from TDVP will ultimately overestimate (for $t>350$)  the actual value if $\chi$ is not sufficient, a behavior similar to that observed for  $L=26$ case mentioned in the previous section.
 
\begin{figure}
  \includegraphics[width=\columnwidth]{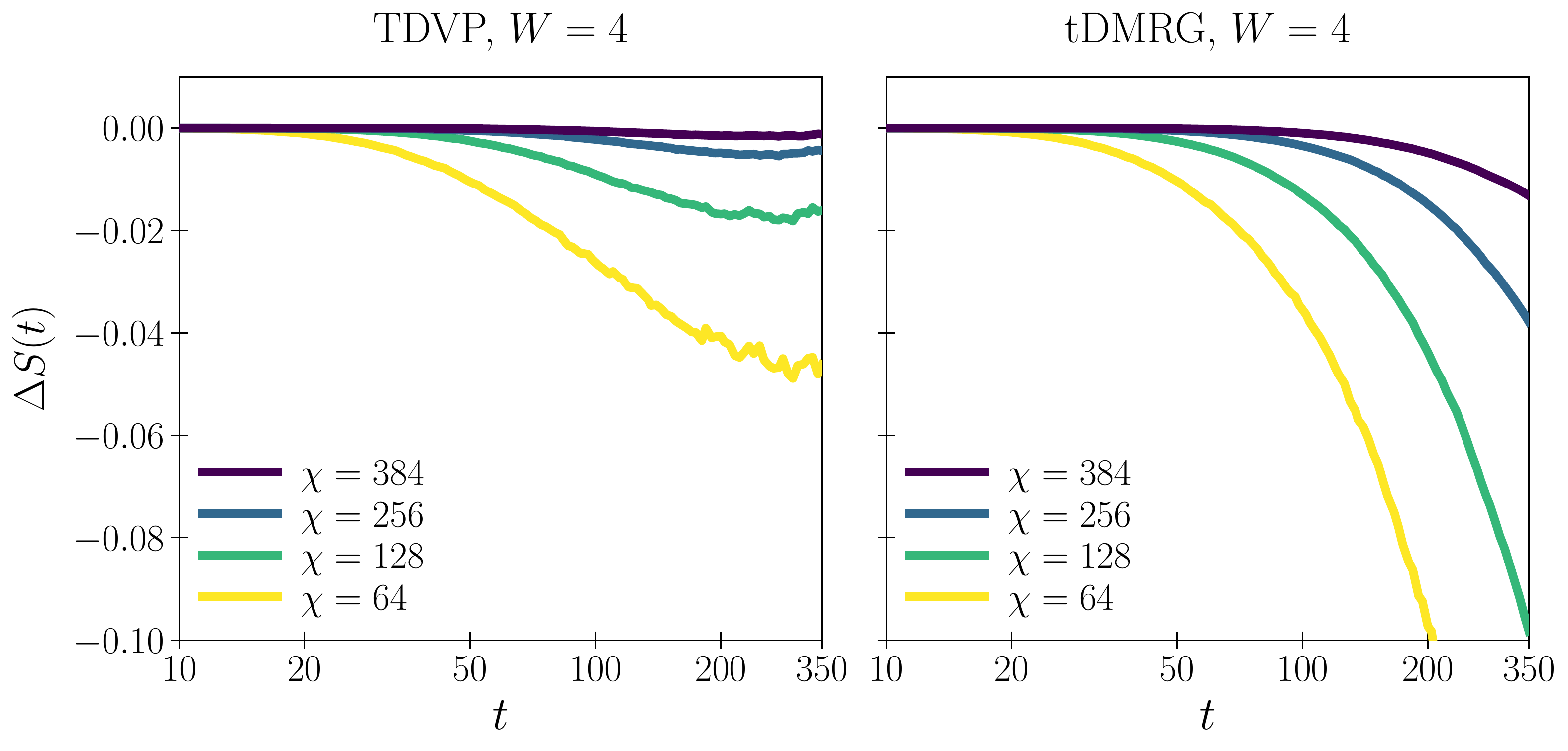}
  \caption{The entanglement entropy error for $L=50$ and $W=4$ corresponding to imbalance depicted in Fig.~\ref{fig50_4imb} for TDVP (top panel) and tDMRG(bottom panel). We plot the difference between the half-chain entanglement entropy for a given $\chi$ (as indicated in the figure) and the corresponding entropy obtained for $\chi=512$. The difference is negligible for TDVP suggesting a convergence of TDVP results in the time window indicated. }
 \label{fig50_4ent}
 \end{figure}
 
\begin{figure}
  \includegraphics[width=\columnwidth]{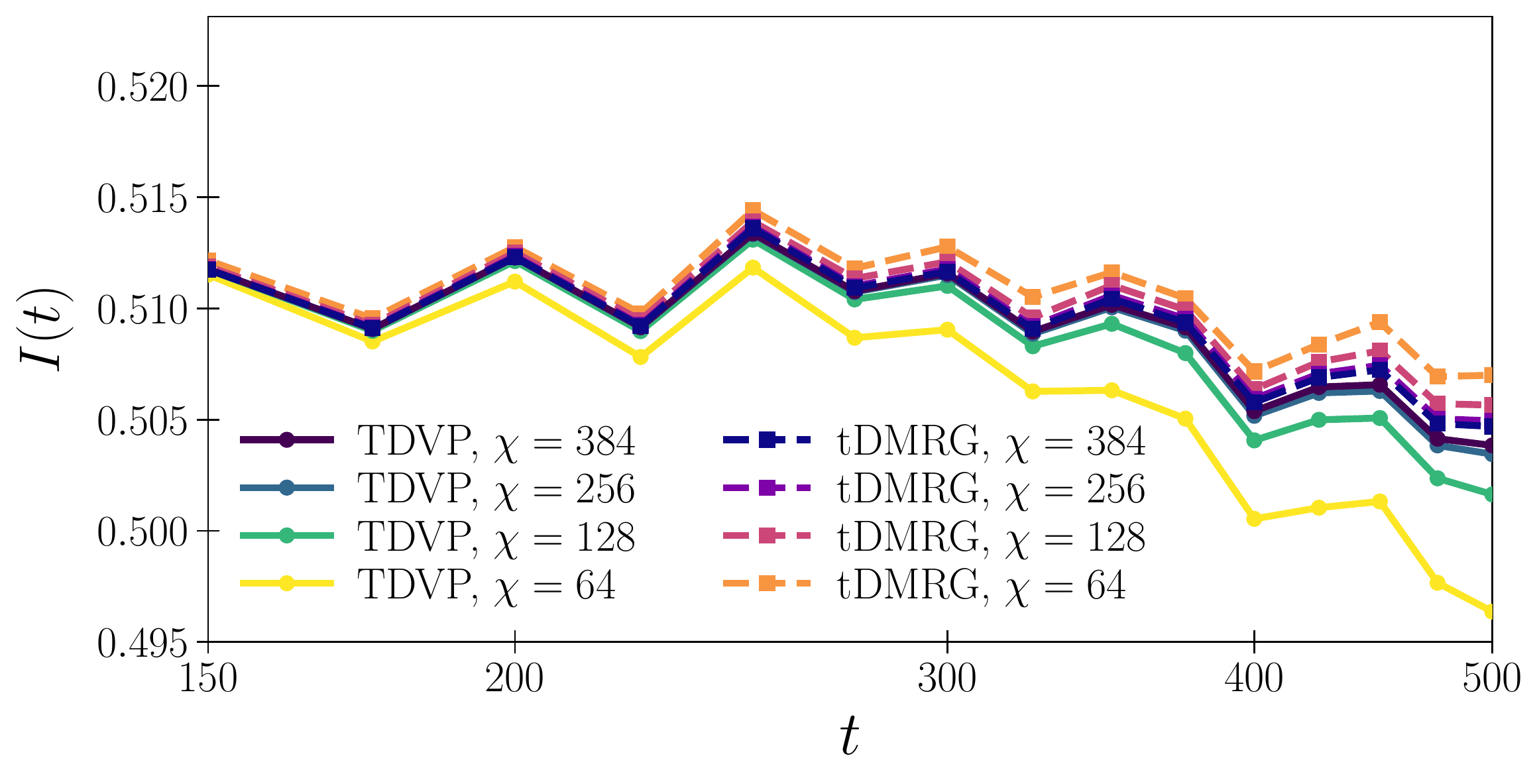}
  \caption{Same as Fig.~\ref{fig50_4imb} but  for $W=5$. To facilitate comparison with earlier $W=4$ case -- compare Fig.~\ref{fig50_4imb} - the similar vertical scale is used. Clearly the convergence is faster for tDMRG simulations. }
 \label{fig50_5imb}
 \end{figure} 
 Finally let us consider even stronger $W=5$ disorder -- Fig.~\ref{fig50_5imb}. Now the convergence of tDMRG is 
 clearly much faster than that of TDVP. All curves are markedly more horizontal indicating a much slower (if any) decay. Note that to facilitate comparison of different $\chi$ values only equally spaced point in times (every $t=25n$ with $n$ being a positive integer)
 are plotted to remove rapid oscillations with a magnitude exceeding the difference between different signals.
 
\section{Analysis of the crossover and its properties}
\label{sec:crit}

The  convergence examples studied in the previous section show significant fluctuations due to a relatively small number $N=200$ of realizations of disorder. Still they are sufficient to draw some preliminary conclusions. Firstly, in the interesting interval of disorder values, $\chi=64$ or even $\chi=128$, as used e.g. in \cite{Doggen18}, lead to unconverged results  affecting
the shape and the decay of imbalance curves. Only for $W>4.5$ such  small values of $\chi$ seem to produce results one could rely on. We decided to compromise on  $\chi=384$ data as a proper choice
for estimating the imbalance decay, for which we have reasonably reliable data upto $t=500$. Still, we may not be, as exemplified in detail above, be convinced about the convergence for $W<4$ even for shorter times. For better statistics from now on we use 400 realizations of disorder for $L=50$, unless otherwise stated.

As discussed earlier
\cite{Luitz16,Luschen17}, it is suggested that the imbalance decays as a power law on a delocalized side of 
ergodic-MBL crossover. However, there exists, as far as we know, no real theoretical foundation for this conjecture. Small system sizes at short times (up to $t=100$) were fitted in \cite{Luitz16} by a nine parameters formula matching also rapid oscillations at very short times with a power-law component, The experimental data \cite{Luschen17} were fitted by the power law decay, $t^{-\beta}$, as well as (see supplementary material of \cite{Luschen17}) by the exponential decay. In the previous study \cite{Doggen18} power law fits in $t\in [50,100]$
interval were analyzed.

\begin{figure}
  \includegraphics[width=0.9\columnwidth]{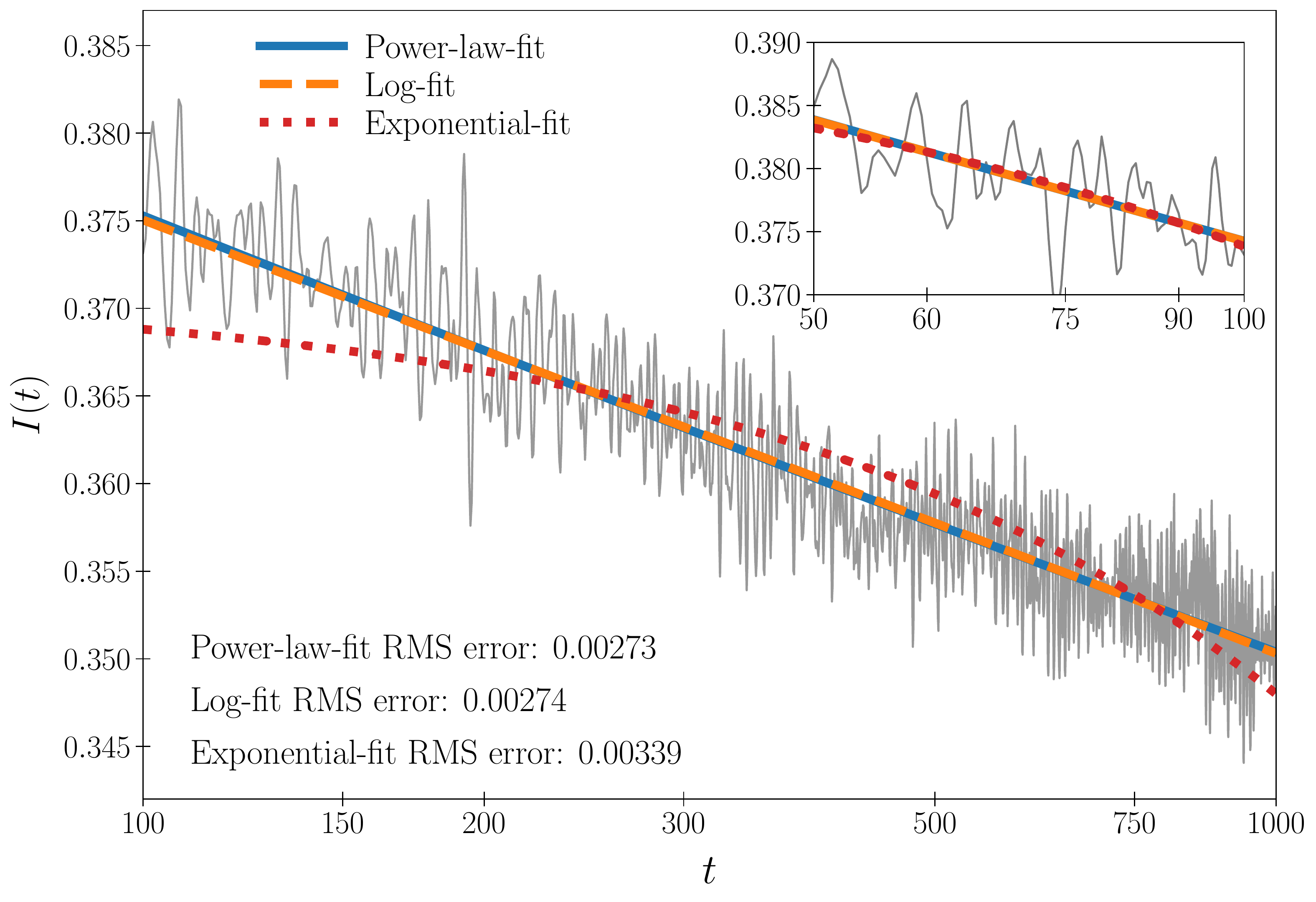}
  \caption{The imbalance for 200 realizations of disorder with $W=3.5$ for $L=26$ as obtained from time-evolution by Chebyshev expansion. The main
  panel shows the decay for long times, up to $t=1000$. Both the power law and logarithmic
  decay match the data with a comparable precision. Exponential decay is by comparison ruled out. The inset shows the data for smaller times where prominent oscillations may affect the fit.}
 \label{fit26}
\end{figure}
The difficulty encountered by fitting the data is exemplified in Fig.~\ref{fit26}. To leave out the convergence problems for a moment, we present the exact data for $L=26$ with 200 disorder realizations. The inset shows the fits of different time-dependencies of the decay
in   $t\in [50,100]$ interval. 
We use the following fitting formulae:
\begin{align}
 f_e(t)&= a_1\exp(-a_2t) {\mathrm{ \qquad exponential\ fit}\qquad} \nonumber \\
 f_p(t)&= a_1/t^{a_2} {\mathrm{\quad \qquad\qquad power-law\ fit}   \qquad } \nonumber\\
 f_l(t)&= a_1 + a_2\ln(t) {\mathrm{\ \qquad logarithmic\ fit.}\qquad}\nonumber
\end{align}
Bearing in mind the existence of significant oscillations in the data for short times, all three functional dependencies assumed (logarithmic, exponential and power law) fit  the average decay reasonably well. 
The main panel shows the decay on a large time scale (available for Chebyshev propagation) where clearly the exponential decay is ruled out but both power law and logarithmic fits are practically indistinguishable. This clearly shows the ambiguity in extracting reasonable critical disorder value by curve-fitting in such small time-windows (even when $t$ is considered up to 1000).

Such large times are practically unreachable with TDVP and tTDMRG due to convergence problems as shown in the previous Section. Therefore we have to make a reasonable compromise, and we settle for $t\in [100,200]$ interval. This increases twice the time interval as compared to \cite{Doggen18} study, makes the role of oscillations weaker (they obviously affect the fit as the maxima and minima shift with the disorder strength), and produces
almost satisfactory convergence in tDMRG and TDVP simulations. From now on we use the power law fit for a better comparison with earlier works.

In fact, the discrepancies observed from not fully converged tDMRG and TDVP data may be quite informative.  MBL is known to be characterized by area-law-entangled eigenvectors \cite{Bauer13}  and slow logarithmic growth of the entanglement entropy from the initial separable state \cite{Znidaric08, Bardarson12}. One could envision that this property could be used in our simulations, but a casual glance at examples given in the previous section, shows that  entanglement entropy converges slowly with $\chi$, moreover making a distinction between logarithmic and power law growth is difficult on short time intervals (as for imbalance). Still in MBL regime we expect a better, faster convergence with increasing $\chi$ as well as smaller differences between TDVP and tDMRG results.

\begin{figure}
  \includegraphics[width=\columnwidth]{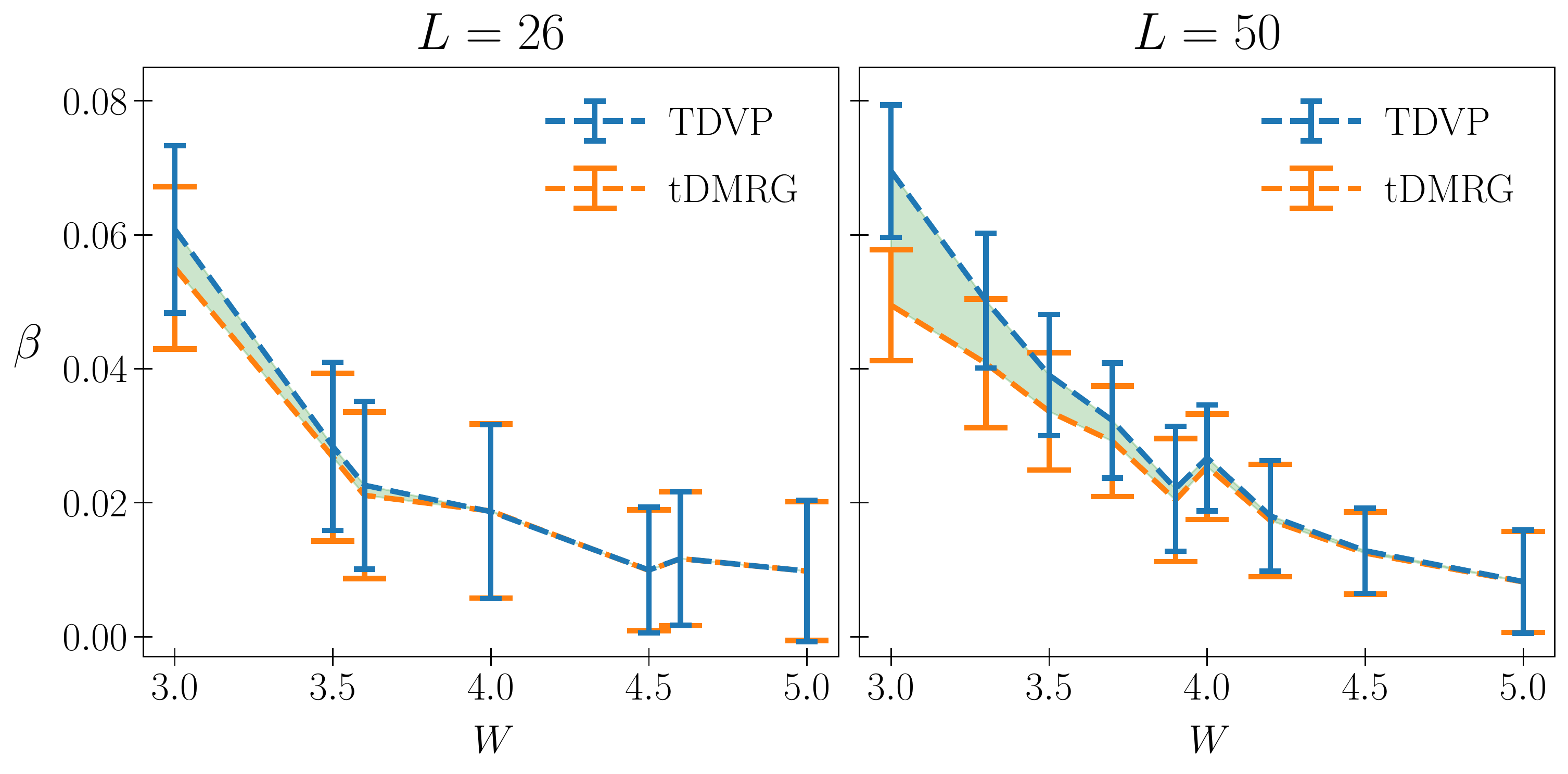}
  \caption{The power law ($t^{-\beta}$) $\beta$ coefficient obtained from fits of {imbalance} obtained from both TDVP and tDMRG  in $t\in [100,200]$ interval for $L=26$ left and $L=50$ right for $\chi=384$ for 200 disorder realizations. The error-bars correspond to $2\sigma$-error obtained from statistical bootstrapping procedure.
  The exact data from Chebyshev time evolution lie within a shaded area for $L=26$. Results of both methods practically coincide for $W>4$ for both $L=26$ and  $L=50$.}
 \label{tdvptebd}
\end{figure}
We now turn to analysis of imbalance decay.
Comparison of power law fits for TDVP and tDMRG are shown in Fig.~\ref{tdvptebd} for 200 disorder realizations. Shaded regions are discrepancies appearing in the delocalized regime. The data for $L=26$ are of course more converged than those for $L=50$ with the same assumed bond dimension $\chi=384$. 
However, for $W>4$, the discrepancies  between fits to TDVP and tDMRG data are independent of the system size. These data suggest that a qualitative change in the behavior of the system occurs around $W\approx 4$. 

\begin{figure}
  \includegraphics[width=0.9\columnwidth]{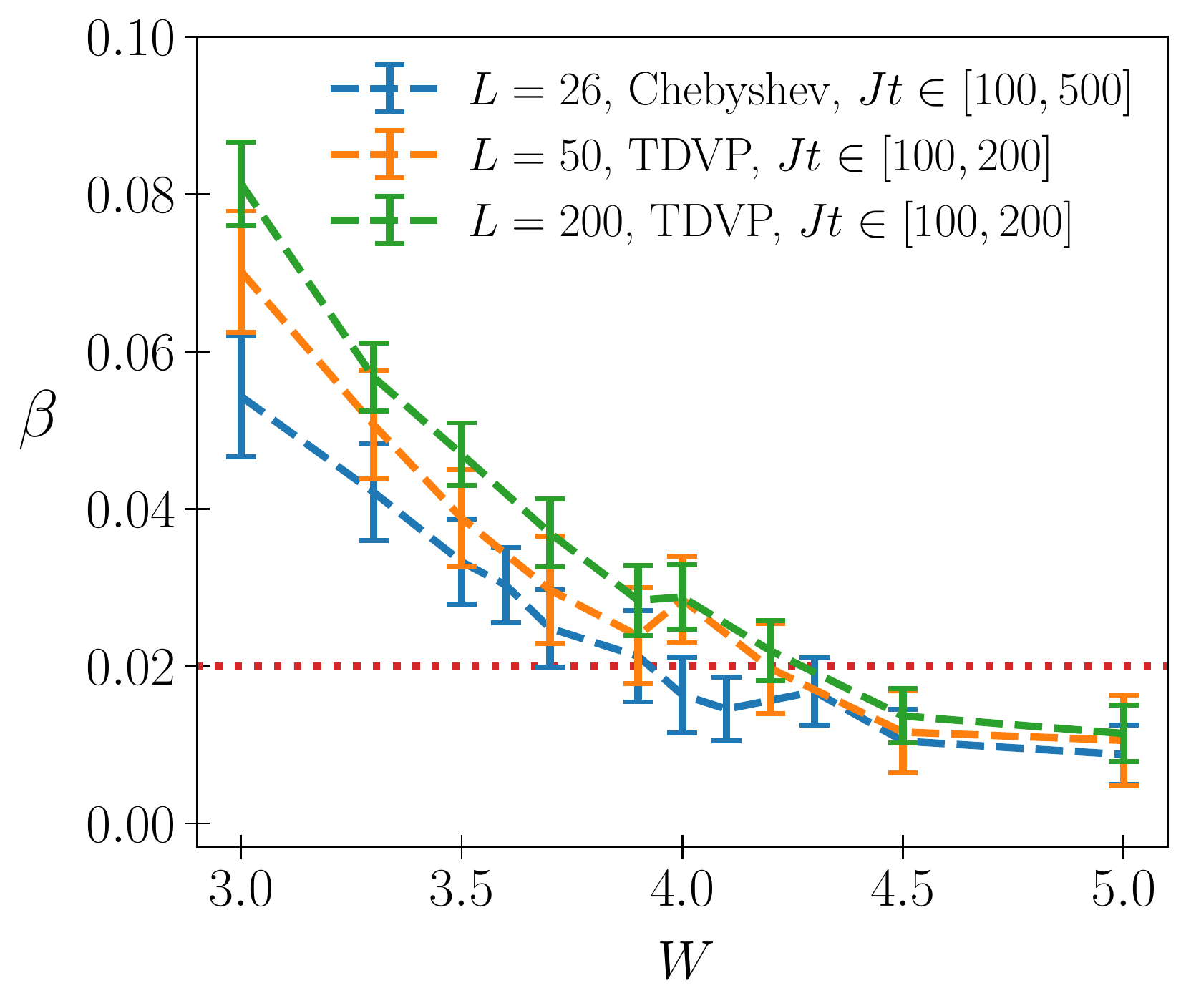}
  \caption{The power law ($t^{-\beta}$) $\beta$ coefficient   obtained from fits of  TDVP data in $t\in [100,200]$ interval for $\chi=384$ and different system sizes. The exact data from Chebyshev time evolution are given for $L=26$ in $t \in [100, 500]$. The error-bars corresponds to $2\sigma$-error obtained from statistical bootstrapping procedure.
 The dashed line at $\beta=0.02$ is our assumed confidence level.}
 \label{tdvp}
\end{figure}
This is further supported by Fig.~\ref{tdvp} showing results of the fits 
where only TDVP data are taken into account. Considering only 400 realizations of disorder
(dictated by the relatively large $\chi$ value in simulations) we deal with quite a noisy data 
as shown in the previous section. We take $\beta=0.02$ as a threshold value that
describes the distinguishability between the power-law decay and the stationary long time behavior. This threshold value roughly corresponds to $4\sigma$-errors of our data by statistical bootstrapping procedure at $W = 4$ for $L=50$, and is also consistent with $W_c$ estimate via level-spacing distribution for smaller systems ($L=20$).
A small shift of the fitted curves with the system size is observed, but nevertheless, the
curves for $L=50$ and $L=200$ practically 
coincide at larger disorder strengths yielding the estimate of the 
critical disorder value around $W \approx 4.2$. For comparison, we 
also fit $L=26$ data obtained from Chebyshev method in a bigger time window $t \in [100, 500]$, as 
convergence problem for longer time can be ruled out in this situation.  Like $L=50$ and $200$, the 
$L=26$ curve also shows a qualitative change around $W \approx 4$. 
Both the weak dependence on the system size and the value of the critical disorder estimate differ
significantly from the conclusions of \cite{Doggen18}. The main difference 
in the value of critical disorder strength, however, originates from the different
``$\beta$ cutoff'' assumed to be at $\beta \approx 0.01$ in \cite{Doggen18}, which does actually shift the
apparent value of $W_c$  close to 5 even for smaller system-sizes like $L=26$.
We also note that the critical disorder value $W\approx 4.2$ obtained in our analysis 
is in good correspondence with results of \cite{Devakul15} that suggest
critical disorder strength to be about $W=4.5(1)$ in the middle of the spectrum.

Taking all the arguments, and in particular the certain arbitrariness of the threshold $\beta$ choice, we give rather large error for our estimate of the critical
disorder as ${W_c=4.2\pm 0.3}$.

\begin{figure*}
 \includegraphics[width=0.95\linewidth]{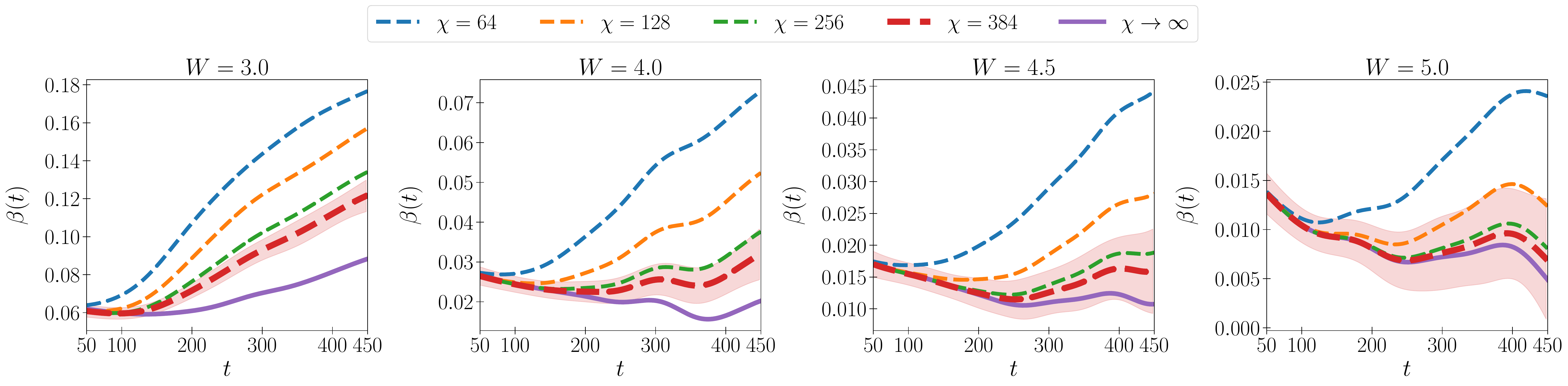}
  \caption{Flowing power-law exponent $\beta(t)$ as derived for different disorder values (indicated in the figure) from the TDVP data for $L=50$ and 800 disorder realizations. Here we also extrapolate the data to $\chi \rightarrow \infty$ to estimate the behavior of $\beta(t)$ at infinite bond dimension.
 Observe a strong dependence on $\chi$ indicating that predictions for small $\chi$ are misleading.
  For $W=4.5$ a decrement in $\beta(t)$ values indicating localization is observed. Shaded areas indicate $2 \sigma$-errors of $\chi=384$ data from bootstrapping procedure.}
 \label{fig:flowing}
 \end{figure*}
 
Let us come back to the analysis of the imbalance decay and make a flowing-$\beta$ analysis of the decay following the procedure described in \cite{Doggen19} for a quasiperiodic disorder. Explicitly, first we assume a window of times $\tau \in [30, 500]$. Then, for each time $t$,
we fit $I(\tau)$ with $\tau^{-\beta(t)}$ where the imbalance data are weighted according to a Gaussian of width $\tilde{\sigma}$, $\sim \exp\left[- \frac{(t-\tau)^2} {2 \tilde{\sigma}}\right]$,
 where $\tilde{\sigma}$ is chosen to be sufficiently large to remove oscillatory (as well as fluctuating) behavior of the data. In our case, we choose $\tilde{\sigma} = 60$.
In localized regime, where the imbalance should saturate eventually, $\beta(t)$ should show a decrement with $t$ in a moderate time window. In Fig. \ref{fig:flowing}, we show the time dependence of $\beta(t)$ derived from the TDVP data of $L=50$ for different disorder strengths and different values of $\chi$.
Here, we used 800 realizations of disorder to minimize the statistical errors in estimating $\beta(t)$.
We observe a strong dependence on $\chi$ indicating that data with low $\chi$ should be taken with caution. In these regards, we also perform $\chi \rightarrow \infty$ extrapolation to estimate the `converged' behavior of $\beta(t)$ at infinite bond dimension.
For $W=3$, the flowing-beta increases with time indicating that the decay of imbalance is faster than a power-law. For $W=4.5$, while small $\chi$ suggests delocalization, sufficiently large $\chi$ values indicate that $W=4.5$ is above $W_c$, as $\beta(t)$ starts to decrease with time. Such decrement of $\beta(t)$ is more prominent for $W=5$.
This analysis of flowing-$\beta$ is indeed in parity with the earlier estimation of critical $W_c = 4.2 \pm 0.3$.

\begin{figure}
  \includegraphics[width=0.9\columnwidth]{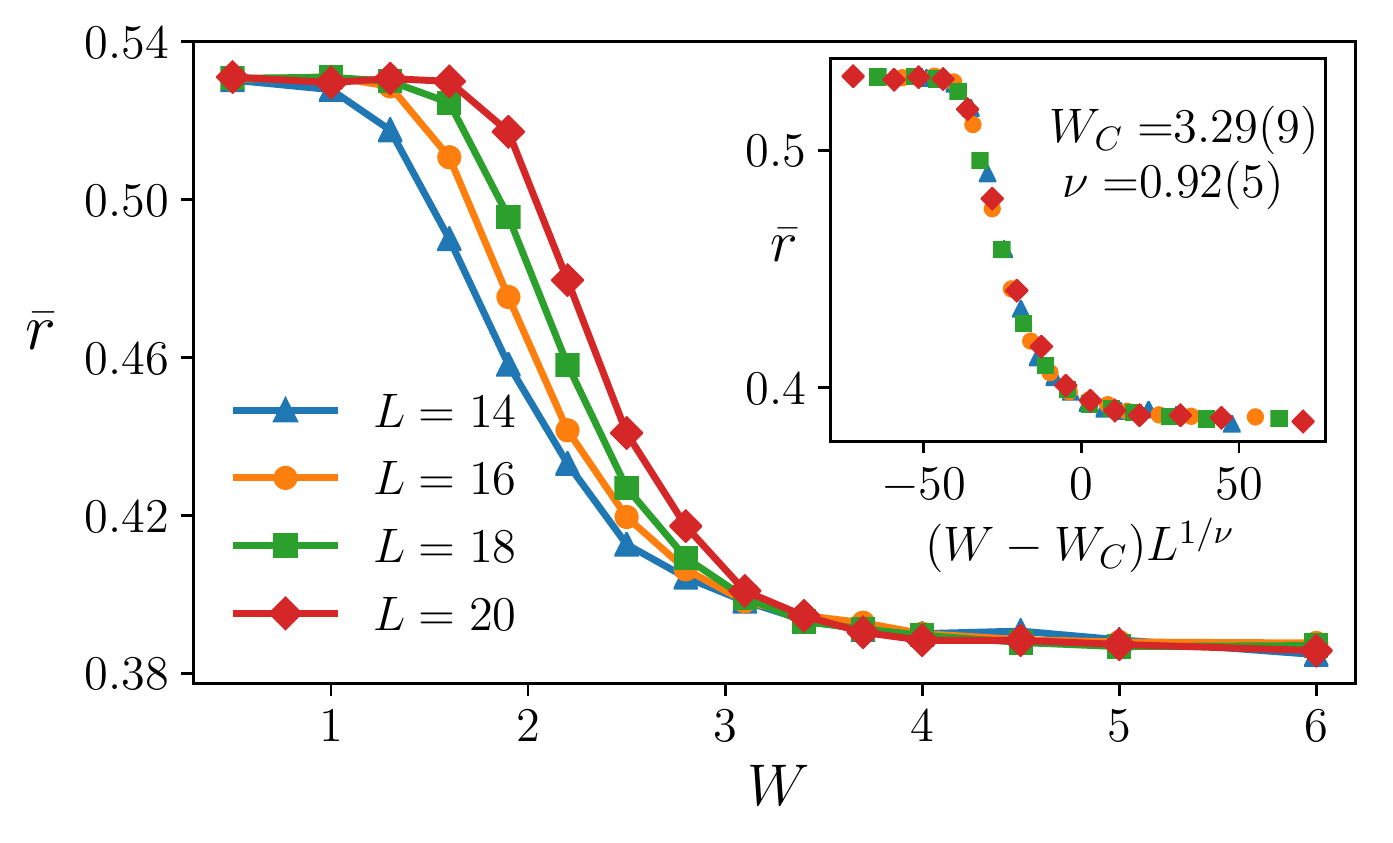}
  \caption{The average level spacing ratio $\overline r$ 
  as a function of disorder strength $W$ for various system sizes $L$ with open boundary condistions. The inset shows collapse 
  of the data upon rescaling of disorder strength according to $W \rightarrow (W-W_C)L^{1/\nu}$. The obtained value of critical disorder strength $W_C=3.29(9)$ is markedly smaller than value 
  found in \cite{Luitz15} for the system with periodic boundary conditions.  }
 \label{gap}
 \end{figure}

In \cite{Doggen18} the critical disorder value was compared with $W_c\approx 3.7$ found in \cite{Luitz15}
on the basis of finite size scaling of the average level spacing ratio $\overline r$ obtained for system sizes 
$L=14,..., 22$. The analysis of \cite{Luitz15} was conducted for a system with periodic 
boundary conditions while time evolution studies are carried out with open boundary conditions (OBC). 
Surprisingly, the choice of boundary conditions strongly affects critical disorder strength value
obtained from finite size scaling of data from level statistics. 
To demonstrate this, we find $500$ eigenvalues $E_i$ from the middle of spectrum of 
system with sizes $L=14,16,18,20$ with shift-and-invert
technique \cite{Pietracaprina18}, calculate the level spacing ratios 
$r_i=\frac{\min \{E_{i+2}-E_{i+1},E_{i+1}-E_{i} \} }{\max\{E_{i+2}-E_{i+1},E_{i+1}-E_{i}\}} $ 
and average it within given disorder 
sample and, subsequently, over $2000$ ($500$) disorder realization for $L=14,16,18$ ($L=20$) 
to obtain
the average level spacing ratio $\overline r$. The results are 
shown in Fig.~\ref{gap}. A clear crossover between $\overline r=0.5307(1)$ in the ergodic regime 
and $\overline r = 0.3863(1)$ in the MBL regime is visible. Rescaling the disorder strength according
to $W \rightarrow (W-W_C)L^{1/\nu}$ leads to a collapse of the data yielding value of critical disorder strength 
 $W_c=3.29(9)$ (and $\nu=0.92(5)$). Discrepancy of the critical value of disorder strength for system with OBC
and the result of \cite{Luitz15} for periodic boundary conditions demonstrates the significance of the edges of the system which 
should be negligible if $W_c$ obtained in this way was truly a thermodynamic limit quantity.
 
Moreover, there is an obvious disagreement between $W_c=3.29(9)$  obtained from finite-size scaling of $\overline r(W)$
and from time evolution of imbalance $W\approx4.2$. We believe that there is no reason to expect 
the critical disorder estimates obtained by these two approaches to coincide. 
 The level spacing ratio $r_n$ contains information about properties of the system at energy scale of 
 single level spacing. The corresponding time scale (Heisenberg time) is exponentially 
 large is system size $L$. Features of level statistics at scales of few and many level spacings 
  \cite{Sierant19b} correspond to shorter time scales and can be  
 observed in time dynamics \cite{Schiulaz19}. Nevertheless, the involved 
  time scales are still exponentially large in system size $L$.
However, the analysis of time-dependence of imbalance necessarily addresses
finite time intervals comparable to experiments in cold atoms and long time
scales are not reachable by present state-of-the-art tools used by us. From 
that perspective, the  time evolution study of ergodic-MBL transition, even 
of very large systems, may only suggest ergodic or localized behavior at some intermediate disorder strengths
whereas the 
precise location of the transition 
point between the two phases can be pinpointed only in limit of infinite times.

\section{Conclusions outlooks}
\label{sec:outlooks}

We have discussed two distinct but related issues. On the technical side, we compare the performance of two
popular schemes for simulation of time-evolution in large  one-dimensional disordered systems --  tDMRG (as a variant of
TEBD) and TDVP. While TDVP outperforms tDMRG for delocalized systems, in the crossover region between delocalized
and MBL side, tDMRG may 
be slightly better. In particular, errors in the entanglement entropy accumulate in tDMRG in 
a controllable and predictable way, while TDVP shows counter-intuitive behavior of overshooting the actual value for smaller auxiliary spaces. 
Interestingly, studies of imbalance indicate 
that tDMRG/TDVP overestimates/underestimates the imbalance at longer times, so joining the information from both algorithms 
allows to estimate the proper imbalance behavior as a function of time.

Our study, extended to longer times, larger system and, importantly much larger $\chi$
show, that at the assumed level of accuracy, the dependence of disordered Heisenberg spin chain on the system 
size is rather weak, contrary to earlier findings \cite{Doggen18}.
While we provide arguments towards the estimate of the critical disorder value at $W_c\approx 4.2$, we must
stress that this estimate is limited to the time interval (up to $t=200$, or $t=500$ for flowing-$\beta$ analysis) - extending this time to much 
larger values is beyond the current computational power - except for systems very deep in the MBL regime. 
With this restriction, our study, especially the flowing-$\beta$ analysis, strongly suggest a saturation of the imbalance in long time limit for large system sizes contradicting the claims of \cite{Suntais19} which questions the MBL existence in the thermodynamic limit. 
For further discussions regarding this interesting issue see \cite{Abanin19, Sierant19c, Panda19}.

\acknowledgments
We thank Marek M. Rams, E. Miles Stoundenmire, and Matthew Fishman for helpful suggestions 
regarding the implementations of the TDVP algorithm.
The MPS-based techniques have been implemented using
ITensor library v2 (\url{https://itensor.org}). Support of the Polish National Science Centre via
grants Quantera 2017/25/Z/ST2/03029 (T.C.) and Opus 2015/19/B/ST2/01028 (P.S. and J.Z.)  is acknowledged.
P.S. thanks Polish National Science Centre for an additional support via Etiuda programme 2018/28/T/ST2/00401. 
The partial support by  PL-Grid Infrastructure is also acknowledged. 

\bibliographystyle{apsrev4-1}
%\bibliography{ref0619}
%merlin.mbs apsrev4-1.bst 2010-07-25 4.21a (PWD, AO, DPC) hacked
%Control: key (0)
%Control: author (72) initials jnrlst
%Control: editor formatted (1) identically to author
%Control: production of article title (-1) disabled
%Control: page (0) single
%Control: year (1) truncated
%Control: production of eprint (0) enabled
%

%\input{rmt02a.bbl}

\end{document}